# CABS-dock web server for the flexible docking of peptides to proteins without prior knowledge of the binding site

Mateusz Kurcinski[§], Michał Jamroz[§], Maciej Blaszczyk, Andrzej Kolinski* and Sebastian Kmiecik*

Department of Chemistry, University of Warsaw, Pasteura 1, 02-093 Warsaw

* To whom correspondence should be addressed. Tel: +48 22 822 02 11 ext. 310; Fax: +48 22 822 02 11 ext. 320; Email: sekmi@chem.uw.edu.pl. Correspondence may also be addressed to Tel: +48 22 822 02 11 ext. 320; Fax: +48 822 02 11 ext. 320; Email: kolinski@chem.uw.edu.pl

[§]The authors wish it to be known that in their opinion the first two authors should be regarded as joint First Authors

**ABSTRACT**

Protein-peptide interactions play a key role in cell functions. Their structural characterization, though challenging, is important for the discovery of new drugs. The CABS-dock web server provides an interface for modeling protein-peptide interactions using a highly efficient protocol for the flexible docking of peptides to proteins. While other docking algorithms require pre-defined localization of the binding site, CABS-dock doesn't require such knowledge. Given a protein receptor structure and a peptide sequence (and starting from random conformations and positions of the peptide), CABS-dock performs simulation search for the binding site allowing for full flexibility of the peptide and small fluctuations of the receptor backbone. This protocol was extensively tested over the largest dataset of non-redundant protein-peptide interactions available to date (including bound and unbound docking cases). For over 80% of bound and unbound dataset cases, we obtained models with high or medium accuracy (sufficient for practical applications). Additionally, as optional features, CABS-dock can exclude user-selected binding modes from docking search or to increase the level of flexibility for chosen receptor fragments. CABS-dock is freely available as a web server at http://biocomp.chem.uw.edu.pl/CABSdock

**INTRODUCTION**

Although protein-peptide interactions play key roles in cell functions, relatively little is known about the structural details of these complexes. The highly dynamic and transient nature of peptide binding makes experimental investigation difficult. Thus, computer aided support, like molecular docking, is needed. The major problem in molecular docking is the treatment of protein flexibility (1,2). Namely, current algorithms are not efficient enough to handle the high conformational fluctuations of peptides. An additional difficulty comes from the necessity of simultaneous dealing with the receptor's flexibility, which, even if small, is extremely costly for most of the computational models.

Computational protein-peptide docking is usually divided into three consecutive steps realized by separate protocols (3): (1) prediction of the binding site(s) on the receptor structure, (2) initial modeling of the peptide backbone in the binding site(s), and finally (3) refinement of the initial protein-peptide complexes to high resolution. The CABS-dock method, presented in this paper, is an attempt to unify all these three steps into single, highly efficient simulation of coupled folding and binding of the peptide to the flexible receptor structure. To our knowledge, such an approach to docking, without prior knowledge of the binding site, has been successful so far only when applied to very short peptides (2 – 4 amino acids) (4). We tested performance of the CABS-dock protocol on a wide benchmark set of protein-peptide complexes with peptides of 5 – 15 amino acids. The benchmark contains 103 bound and 68 unbound protein receptor cases (determined experimentally in complex with a peptide and without a peptide, respectively).

**MATERIAL AND METHODS**

**Previous applications and background**

The CABS-dock docking protocol was developed and validated during the following simulation studies: mechanism of folding and binding of an intrinsically disordered peptide (5); docking antigen-mimicking peptides to an antibody (6); and docking peptide co-activators to nuclear receptors (7,8). These studies showed that the method is able to predict complex arrangements close to the native structure. Importantly, in all the validation tests mentioned above, peptides were allowed to be fully flexible and no information about the binding site or peptide conformation was used.

The CABS-dock protocol is a multiscale modeling procedure based on the coarse-grained CABS protein model. The CABS model has been designed to provide significant efficiency in the treatment of protein conformational changes, while preserving high local accuracy (enabling seamless reconstruction to all-atom representation). In the CABS model, each amino acid is represented by up to four interaction centers, simulation dynamics is controlled by the Monte Carlo scheme, and the force field is based on statistical potentials (details have been described elsewhere (9)). Additionally to the aforementioned protein docking studies, we've demonstrated that the CABS protein model enables reliable simulations of protein dynamics: long-term folding mechanisms (10,11), and short-term fluctuations close to the native state (12,13). CABS has also been successfully used in protein structure prediction exercises, showing exceptional performance especially in blind predictions of short globular proteins (14) and large protein fragments (15,16). Altogether, these studies demonstrate the validity of the CABS interaction model and sampling scheme in simulations of simultaneous folding and binding, such as performed in the CABS-dock protocol.

**Protocol overview**

The CABS-dock protocol consists of the following steps (presented also in **Figure 1** and **Figure S1** flow-chart):

(1) Generating random structures. Random structures of the peptide are generated and randomly placed on the surface of the sphere centered at the receptor's geometrical center (the radius of the sphere is the receptor dimension in the longest direction + 20 angstroms).

(2) Simulation of binding and docking. The CABS-dock procedure utilizes Replica Exchange Monte Carlo dynamics with 10 replicas uniformly spread on the temperature scale. Additionally the temperatures of the replicas constantly decrease as the simulation proceeds to end on the bottom of the energy minima. On output the procedure produces 10 trajectories (one for each replica), each consisting of 1000 time-stamped simulation snapshots for a combined total of 10000 models. During the simulation, the receptor molecule is kept in near-native conformations by a set of distance restraints binding pairs of C-alpha atoms. The restraints are selected from the distance map calculated on the input structure based on the following conditions: only C-alpha atoms located within a 5-15 Å range from each other are restrained; the minimum sequence gap between restrained residues is set to 5; violation of the restraint by less than 1 Å is not penalized; beyond that the energetic penalty increases linearly. If the user marks some of the residues as semi-flexible or fully flexible, the slope of the penalty is halved or set to 0, respectively, for all restraints assigned to the marked residues.

(3) Selection of the final models is a two-step procedure:

- Initial filtering. From each of the 10 trajectories, all unbound states are excluded and next 100 lowest binding energy models are selected (or less if a trajectory contains less than 100 bound states, which is rarely the case), for the next step of the procedure.

- K-medoids clustering. Selected models (1000 in total) are clustered together in the k-medoids procedure. Clustering is performed 100 times with different initial medoids and k=10. Ten consensus medoids are selected as the final models.

(4) Reconstruction of the final models. Final models are reconstructed from the C-alpha trace to an all-atom representation and subsequently undergo optimization using Modeller (17) with DOPE statistical potential (18).

**PERFORMANCE**

**Docking without prior knowledge of the binding site**

We have validated the CABS-dock docking protocol against the largest dataset of non-redundant peptide–protein interactions (<70% sequence identity with respect to the receptor protein) available to date. The benchmark was originally created to test the FlexPepDock refinement procedure (3), and subsequently extended (with new unbound cases) in a study testing the HADDOCK algorithm (docking driven by knowledge of the binding site) (19).

We assess the quality of docking models using ligand RMSD (root mean square deviation) as follows:

- High-quality prediction: RMSD < 3 Å

- Medium-quality prediction: 3 Å ≤ RMSD ≤ 5.5 Å

- Low-quality prediction: RMSD > 5.5 Å

The RMSD is calculated on the peptide only, after superimposition of the receptor structures. We've decided to set up an arbitrary cut-off of 5.5 Å (between low and medium accuracy models) on the basis of the work benchmarking the Rosetta FlexPepDock protocol for the refinement of coarse models of protein-peptide complexes (3). In that work, the authors defined an effective "basin of attraction" of 5.5 Å resolution, from which the FlexPepDock protocol is able to reliably recover near-native protein-peptide models (in 91% of bound docking cases). Importantly, low-quality prediction (as defined by the 5.5 Å cut-off) doesn't mean that the obtained models are useless for further refinement. In the FlexPepDock benchmark, starting the refinement from structures of 6.5–7.5 Å resolution resulted in near-native models in 48% of cases of bound docking (3).

In our performance test, we used neither information about the bound peptide structure nor about the binding site (blind prediction test). As shown in **Figure 2**, within the set of resulting models, high or medium accuracy models can be found, both in the entire set of predicted models (all) or in the top selected models (top 10) (for the detailed results of bound and unbound cases see **Table S1** and **Table S2**, respectively). Moreover, the CABS-dock performance for bound cases is on the same level as that for unbound cases (the pairs of bound and unbound counterparts are listed in **Table S3**). This is because, for the majority of benchmark cases, the difference between binding interfaces of bound and unbound protein forms is small (lower than 1 Å (19)). Therefore, such small protein changes are well handled by CABS-dock using the default settings of protein receptor flexibility.

In summary, for over 80% of bound and unbound dataset cases, we obtained models with high or medium accuracy that is sufficient for practical applications (at least for further refinement to higher resolution (3)). The prediction results may qualitatively differ between different prediction runs (due to stochastic nature of the simulation model). The analysis of the accuracy of the predicted models in different prediction runs showed that the modeling cases generally fall into two categories: those with consistent and those with qualitatively distinct predictions (see **Table S1** and **Table S2**). Therefore in ambiguous prediction cases, we suggest running a few independent runs, and analyzing the predictions jointly.

**SERVER DESCRIPTION**

**Input interface and requirements**

The required input includes:

- Protein receptor structure in the PDB format or protein PDB code along with the chain identifier(s), for example: 1RJK:A or 1CE1:HL. The following requirements apply to the input PDB files: single or multi-chain proteins are accepted (chains must be up to 500 amino acids in length); each residue in the provided PDB file should have a complete set of backbone atoms (i.e.: N, Cα, C and O); side chain atoms may be missing. Non-standard amino acids are automatically changed to their standard counterparts.

- Peptide sequence (in one letter code, standard amino acids only, maximum 30 amino acids in length)

The optional input includes:

- Peptide secondary structure. If not provided, the PSI-PRED method (20) for secondary structure prediction is automatically used. For best results, if the peptide secondary structure was derived experimentally, we suggest providing experimental assignments. The secondary structure should be provided for each residue in a three letter code: H, helix; E, extended state (beta strand); and C, coil (less regular structures). Overpredictions of the regular secondary structure (H or E) are more dangerous for the quality of the results than underpredictions (i.e. for residues with an ambiguous secondary structure prediction assignment, it is better to assign coil (C) than a regular (H or E) structure).

- Project name. Recommended for better organization of users' work: project names appear in the queue page.

- Email address. If provided, the server will send an email notification about job completion.

- Advanced options (described in a separate paragraph below).

**Output interface**

The output interface is organized under the following tabs: 'Project information', 'Docking prediction results', 'Clustering details' and 'Contact maps'. The content of the latter three is briefly described in the following paragraphs.

Under the "Docking prediction results" tab (see the screenshot in **Figure 3 A**) the user may view in 3D and download 10 final models (representatives of 10 structural clusters found in the simulation – for details see Protocol overview in Methods). Additionally, the user may download a compressed archive with 10 final models, cluster models (all models that have been classified in structural clustering to particular clusters) and complete trajectories (in the PDB format and C-alpha representation). The archive also contains the input structure of the receptor and a log file with all information to recreate the simulation.

Under the "Clustering details" tab (see the screenshot in **Figure 3 B**) the user is provided with thorough insight into structural clustering data. An interactive chart shows the composition of clusters of models vs. their trajectory affiliation. By clicking on the points representing models in the chart, the user may view in 3D and download particular models in the PDB format. The tab also contains a table summarizing basic information on clusters, such as density, diversity, etc.

The "Contact maps" tab allows the user to investigate the interaction interface between the peptide and the receptor. An interactive chart shows a contact map between the peptide and the receptor residues. The user may define the value of the contact cutoff distance. The user also has a text list of the contacts shown on the maps.

**Advanced options**

For more advanced prediction runs, users may perform the following operations:

- Excluding binding modes. The user may select receptor residues that are unlikely to interact with the peptide to exclude some binding modes from the results. The user may provide such a list explicitly (available from the main page by checking the "Mark unlikely to bind regions" option) or resubmit an already completed job (available in 'Project information' tab) with marked models (binding modes) to be excluded from future results.

- Allowing for the higher flexibility of selected receptor fragments. This advanced option enables removing distance restraints that keep the receptor structure in a near native conformation (see Protocol overview in Methods section). For each selected residue, the user may choose from two preset settings: moderate or full flexibility. This option is available from the main page by checking the "Mark flexible regions" option.

- Increasing simulation length. The user may increase the default (50) number of Monte Carlo macrocycles to be performed (the maximum number is 200). Increasing this number may be beneficial in more difficult modeling cases (e.g. for large receptor structures or for long peptides of more than 20 residues). However, an alternative way for dealing with such more demanding cases is to run independent simulation runs, and to analyze the results jointly.

**Online documentation**

CABS-dock documentation is available online and can be accessed using the links in the menu at the top of every server page. It contains a description of the method and a tutorial explaining how to access and interpret resulting data. The online documentation is updated on a regular basis according to users' needs or method improvements.

**Server and output data availability**

The CABS-dock server is free and open to all users, and there is no login requirement. After clicking on the submit button, a web link to the results is provided which can be bookmarked and accessed at a later time. Web links to the submitted jobs are also displayed on a queue page (available from the main page), unless the option 'Do not show my job on the results page' (available from the options panel in the main page) is marked. Note that the results will be available for a limited period of time (notification about data storage is displayed at the bottom of the job page).

**Server architecture and run-time**

The CABS-dock website interface and parsers were developed in the Python scripting language, using Flask framework and Jinja2 template engine. The molecular visualization used in the server is executed using 3Dmol.js (21) and JSmol (22). The CABS-dock website runs on the Apache2 and SQLite3 database for user queue storage. The CABS-dock queue is checked every 5 minutes by computational servers and any new jobs are added to the SGE queue. As soon as a job is started on the computational server, job status changes on the CABS-dock website (from 'pending' to 'running').

A Typical CABS-dock run takes about 3 hours. After completion, job results are sent back to the website and job status changes from 'running' to 'done' (or 'error'). Currently, CABS-dock server computations are performed on a Linux supercomputer cluster having about 100 CPU threads.

**SUMMARY**

The CABS-dock protocol has been already successfully used in studies of protein-peptide interactions (5-8). Within this work, we developed an easy-to-use web server interface for the CABS-dock protein-peptide docking protocol. We expect that this web server will be widely applied to new systems as well as utilized as an element of new modeling procedures. The promising CABS-dock extensions include for example: adding a refinement step (19,23), incorporation of experimental data, increasing the flexibility of appropriate receptor fragments (e.g. through predicted restraints (24)).

**SUPPLEMENTARY DATA**

Supplementary Data are available at NAR Online: Figure S1, Table S1, Table S2, Table S3.

**FUNDING**

This work was supported by the Foundation for Polish Science TEAM project [TEAM/2011-7/6] co-financed by the EU European Regional Development Fund operated within the Innovative Economy Operational Program; and     Polish Ministry of Science and Higher Education [IP2014 016573]. Funding for open access charge: Foundation for Polish Science TEAM project [TEAM/2011-7/6].

**TABLE AND FIGURES LEGENDS**

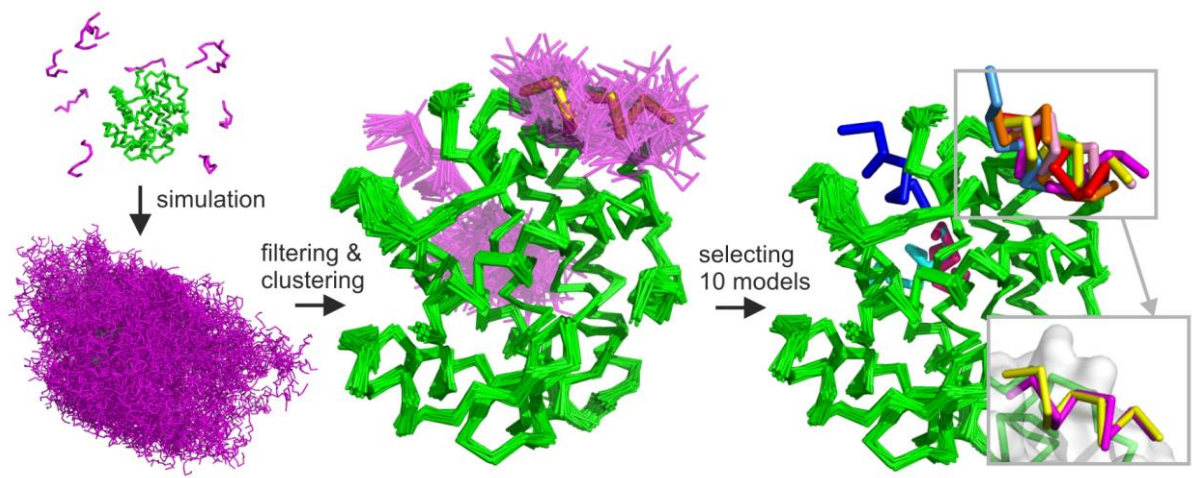

Figure 1. Basic stages of the CABS-dock protocol illustrated on the example benchmark case (PDB ID: 2P1T). The protein receptor is colored in green, modeled peptide conformations in magenta and the reference experimental peptide structure in yellow. The following CABS-dock stages are visualized: (1) simulation start (from random conformations and positions of the peptide); (2) simulation result (a set of 10,000 models); (3) filtering and clustering result (a set of models grouped in similar binding modes and similar peptide conformations); (4) final models (a set of 10 representative models). In the presented benchmark case, 7 of the 10 final models were docked in the native binding site (marked in red rectangle). Among these, the best accuracy model was within 1.37 Å to the native (shown in the right bottom corner superimposed on the native peptide structure). For a more detailed flow-chart of the CABS-dock pipeline, see Figure S1.

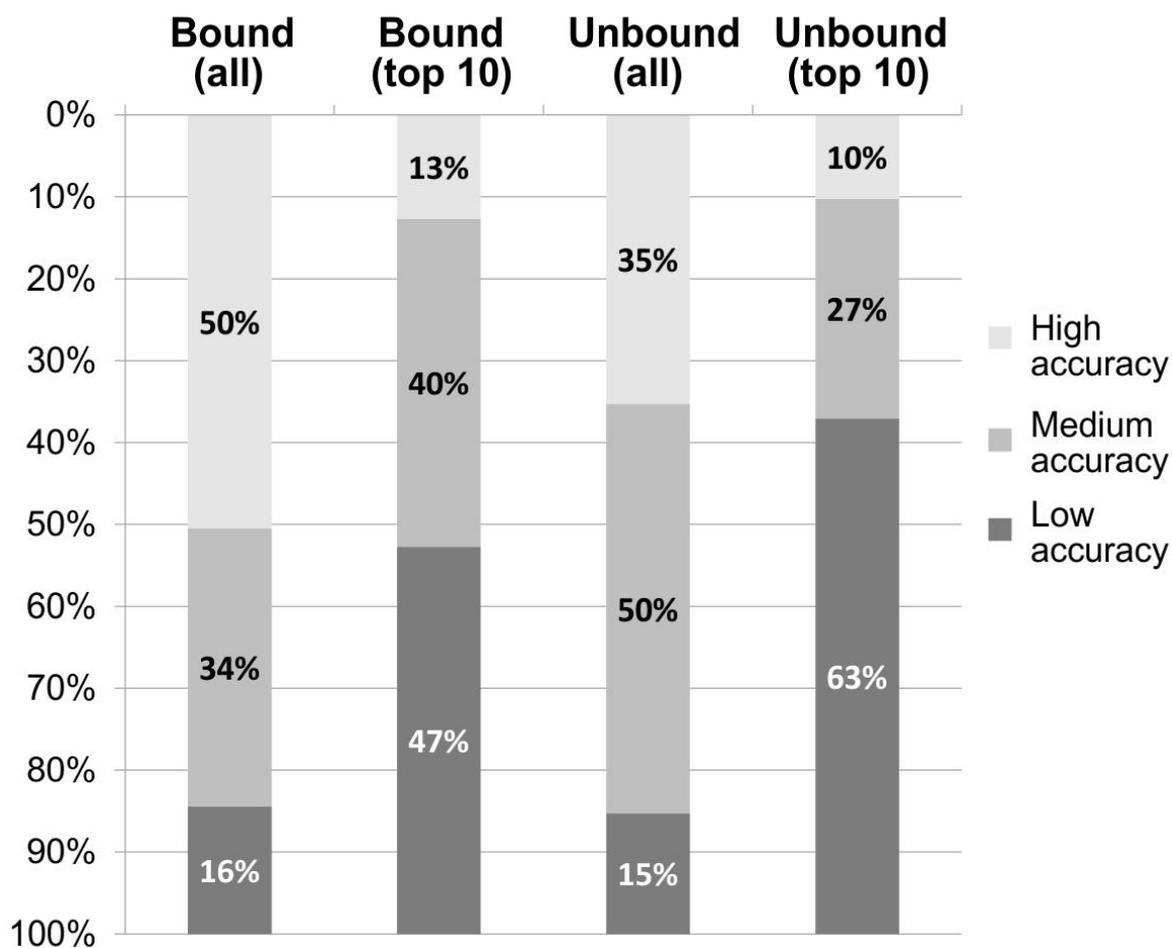

Figure 2. CABS-dock performance summary for 103 bound and 68 unbound benchmark cases. The percentages of high, medium or low accuracy models (quality assessment criteria are given in the text) are reported over all generated models (all: 10,000 models) and top 10 selected models. Detailed results, for each modeled complex and each prediction run, are available in Table S1 (bound docking cases) and Table S2 (unbound docking cases).

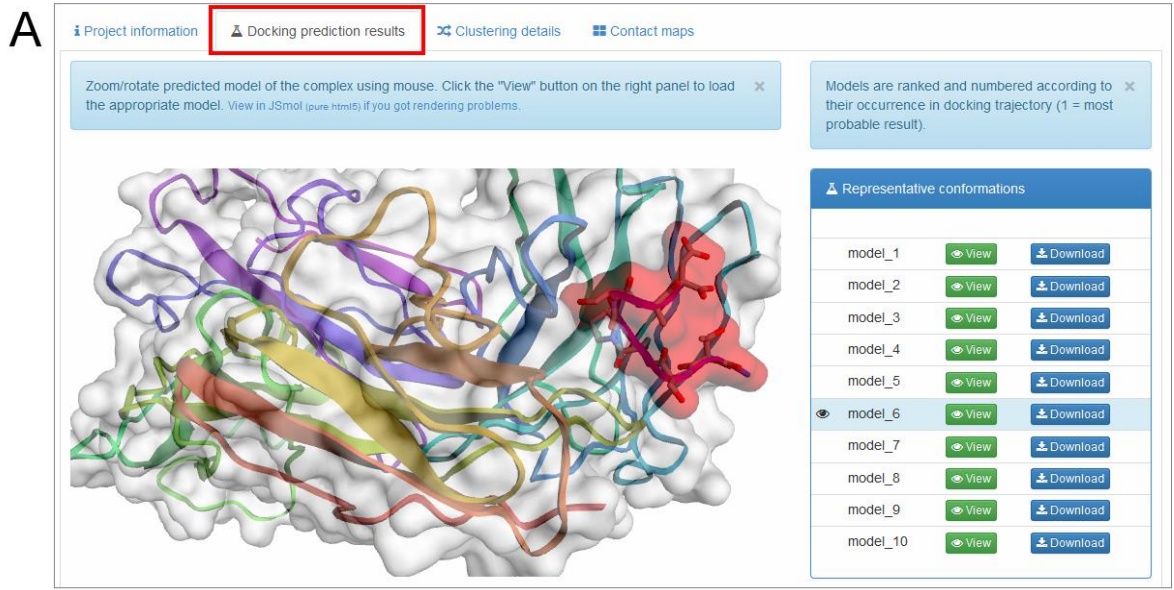

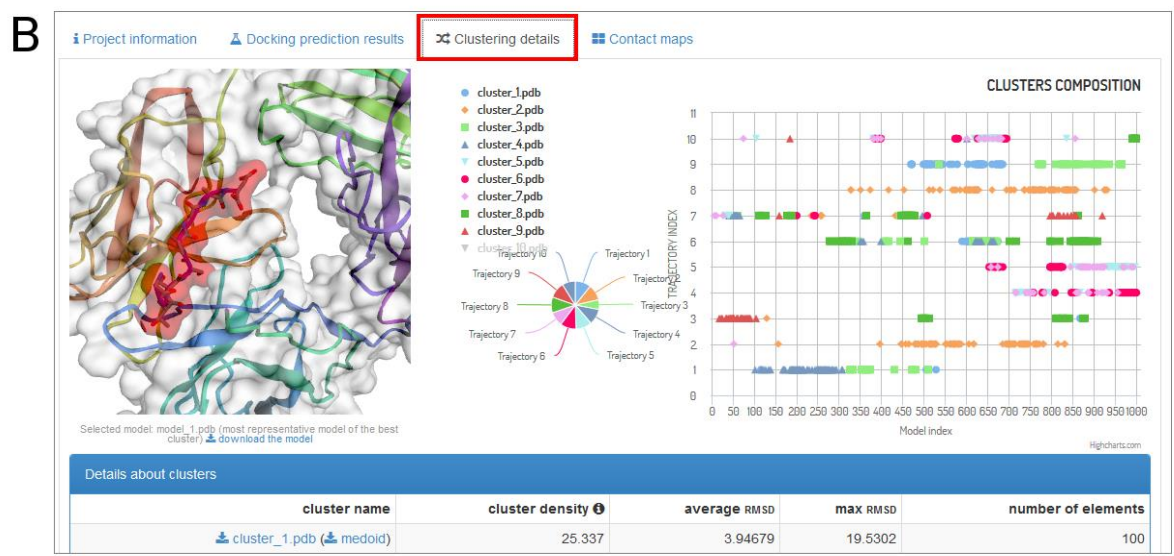

Figure 3. CABS-dock web server screenshots. Example output interface is presented: (A) 'Docking prediction results' tab, and (B) 'Clustering details' tab.

## Supplementary data

**Figure S1. Flow-chart of the CABS-dock protocol.**

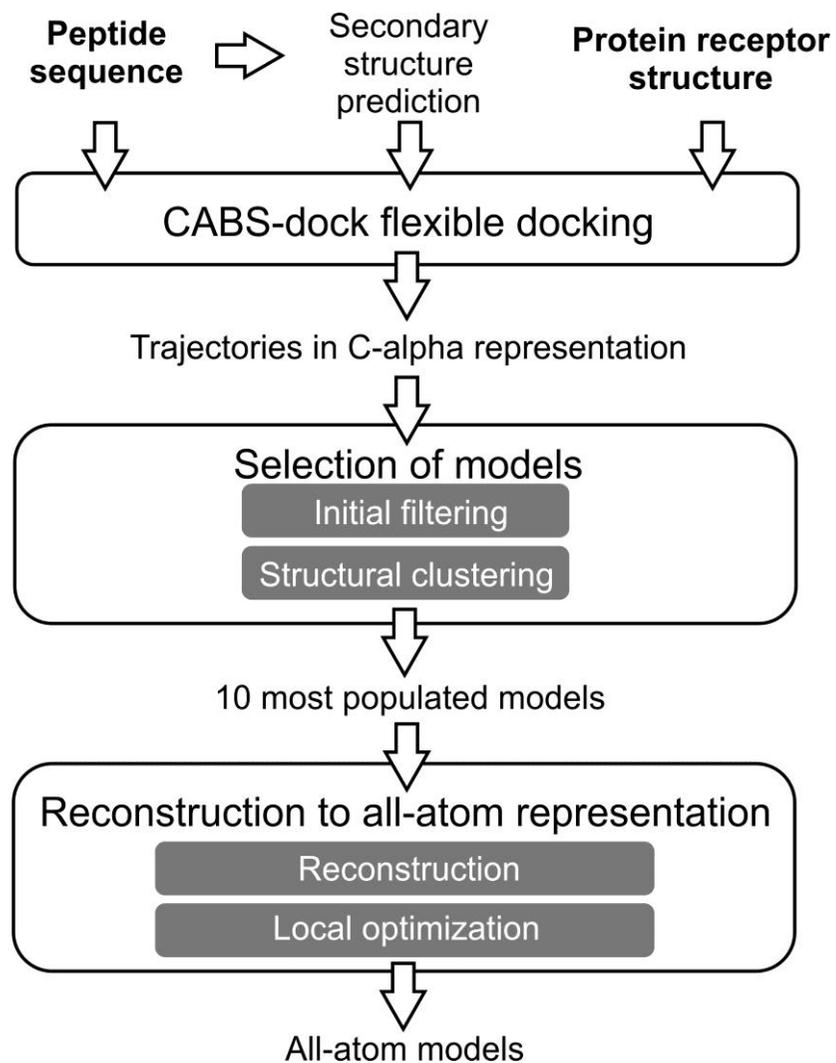

Table S1. CABS-dock performance for 103 bound cases (listed in rows) in 3 independent prediction runs (shown in separate columns). The table shows the lowest ligand-RMSD values (calculated on the peptide only after superimposition of the receptor structures) among: all 10,000 models (all), top 1000 models (top 1k, selected during filtering and clustering procedure), top 100 models (top 100, selected through further clustering) and top 10 final models (top 10). Last column shows the lowest RMSD values obtained in three prediction runs.

| receptor pdb id | receptor length (AA) | peptide length (AA) | Run 1 | | | | Run 2 | | | | Run 3 | | | | best from all runs | | | |
|---|---|---|---|---|---|---|---|---|---|---|---|---|---|---|---|---|---|---|
| | | | all | top 1k | top 100 | top 10 | all | top 1k | top 100 | top 10 | all | top 1k | top 100 | top 10 | all | top 1k | top 100 | top 10 |
| 1awr | 164 | 6 | 1.51 | 2.48 | 2.92 | 2.83 | 1.99 | 1.99 | 2.82 | 5.52 | 1.86 | 1.89 | 2.83 | 3.97 | 1.51 | 1.89 | 2.82 | 2.83 |
| 1ce1 | 431 | 8 | 1.73 | 2.14 | 3.34 | 4.23 | 2.76 | 3.31 | 3.76 | 5.45 | 3.12 | 3.48 | 4.32 | 3.86 | 1.73 | 2.14 | 3.34 | 3.86 |
| 1cka | 57 | 9 | 3.45 | 3.54 | 4.73 | 5.03 | 3.4 | 3.73 | 4.36 | 5.08 | 3.09 | 3.37 | 4.03 | 4.45 | 3.09 | 3.37 | 4.03 | 4.45 |
| 1czy | 168 | 7 | 2.68 | 4.51 | 11.58 | 18.99 | 1.91 | 2.21 | 2.4 | 3.05 | 2.39 | 3.37 | 6.06 | 11.05 | 1.91 | 2.21 | 2.4 | 3.05 |
| 1d4t | 104 | 11 | 4.8 | 5.61 | 7.37 | 8.17 | 1.97 | 1.97 | 2.2 | 2.63 | 2.15 | 2.97 | 3.18 | 3.85 | 1.97 | 1.97 | 2.2 | 2.63 |
| 1ddv | 104 | 6 | 2.52 | 2.94 | 3.97 | 3.82 | 3.24 | 3.88 | 4.75 | 4.07 | 3.08 | 3.53 | 4.7 | 6.92 | 2.52 | 2.94 | 3.97 | 3.82 |
| 1dkx | 219 | 7 | 8.49 | 8.49 | 9.11 | 8.84 | 7.93 | 7.93 | 8.19 | 9.4 | 9.59 | 9.74 | 10.29 | 12.78 | 7.93 | 7.93 | 8.19 | 8.84 |
| 1eg4 | 260 | 13 | 6.73 | 8.61 | 8.97 | 11.08 | 9.57 | 11.38 | 12.26 | 16.18 | 6.57 | 13.12 | 13.39 | 13.7 | 6.57 | 8.61 | 8.97 | 11.08 |
| 1elw | 117 | 8 | 1.96 | 1.96 | 2.53 | 3.8 | 2.18 | 2.94 | 3.3 | 4.87 | 2.56 | 2.56 | 3.38 | 4.57 | 1.96 | 1.96 | 2.53 | 3.8 |
| 1er8 | 330 | 8 | 6.33 | 6.92 | 6.92 | 6.92 | 7.11 | 7.11 | 7.78 | 8.68 | 6.39 | 6.39 | 6.9 | 8.9 | 6.33 | 6.39 | 6.9 | 6.92 |
| 1gyb | 122 | 5 | 4.53 | 8.62 | 11.55 | 16.66 | 4.79 | 7.5 | 14.76 | 15.15 | 4.98 | 7.67 | 10.99 | 14.83 | 4.53 | 7.5 | 10.99 | 14.83 |
| 1h6w | 151 | 10 | 6.1 | 18.63 | 19.07 | 25.35 | 8.22 | 8.22 | 19.27 | 20.32 | 9.46 | 10.2 | 12.11 | 25.23 | 6.1 | 8.22 | 12.11 | 20.32 |
| 1hc9 | 74 | 13 | 6.87 | 10.59 | 12.91 | 14.26 | 6.21 | 6.52 | 7.17 | 7.94 | 6.67 | 11.81 | 13.69 | 14.5 | 6.21 | 6.52 | 7.17 | 7.94 |
| 1i8k | 225 | 10 | 3.6 | 4 | 4.04 | 8.5 | 5.53 | 5.82 | 7.2 | 7.64 | 4.92 | 5.5 | 6.94 | 7.8 | 3.6 | 4 | 4.04 | 7.64 |
| 1iak | 367 | 13 | 3.4 | 4.19 | 4.19 | 5.13 | 5.56 | 8.19 | 10.16 | 18.04 | 3.98 | 3.98 | 5.2 | 5.85 | 3.4 | 3.98 | 4.19 | 5.13 |
| 1ihj | 94 | 5 | 2.58 | 3.71 | 4.42 | 6.99 | 2.7 | 2.72 | 5.29 | 6.5 | 2.17 | 4.84 | 5.67 | 7.03 | 2.17 | 2.72 | 4.42 | 6.5 |
| 1jbu | 239 | 15 | 7.73 | 9.67 | 13.45 | 19.68 | 9.39 | 14.52 | 14.52 | 26.58 | 7.39 | 13.46 | 21.87 | 24.87 | 7.39 | 9.67 | 13.45 | 19.68 |
| 1jd5 | 105 | 8 | 2.85 | 2.85 | 5.88 | 16.12 | 5.31 | 13.06 | 15.47 | 20.07 | 6.04 | 9 | 11.69 | 15.18 | 2.85 | 2.85 | 5.88 | 15.18 |
| 1jwg | 140 | 5 | 2.53 | 3.48 | 4.41 | 5.27 | 2.29 | 2.29 | 3.92 | 6.75 | 2.38 | 2.38 | 4.21 | 5.87 | 2.29 | 2.29 | 3.92 | 5.27 |
| 1kl3 | 120 | 6 | 2.6 | 3.51 | 5.14 | 5.33 | 2.91 | 4.33 | 4.82 | 5.92 | 2.45 | 4.29 | 4.92 | 5.21 | 2.45 | 3.51 | 4.82 | 5.21 |
| 1klu | 369 | 15 | 3.57 | 3.89 | 4.79 | 5.86 | 3.64 | 3.91 | 4.5 | 5.3 | 3.75 | 3.8 | 5.02 | 6.27 | 3.57 | 3.8 | 4.5 | 5.3 |
| 1lvm | 229 | 7 | 6.13 | 7.88 | 17.09 | 13.47 | 5.68 | 9.59 | 12.52 | 14.37 | 5.42 | 6.07 | 14.41 | 15.07 | 5.42 | 6.07 | 12.52 | 13.47 |
| 1mfg | 95 | 9 | 4.62 | 4.86 | 6.78 | 9.49 | 2.99 | 4.6 | 10.1 | 7.37 | 3.71 | 3.71 | 7.25 | 6.78 | 2.99 | 3.71 | 6.78 | 6.78 |
| 1mvu | 333 | 13 | 4.26 | 4.26 | 5.47 | 9.85 | 3.74 | 4.33 | 4.89 | 5.29 | 3.29 | 4.84 | 6.58 | 7.77 | 3.29 | 4.26 | 4.89 | 5.29 |
| 1n12 | 136 | 11 | 8.43 | 8.43 | 9.4 | 12.11 | 8.31 | 8.31 | 8.72 | 10.46 | 9.46 | 9.89 | 12.78 | 14.93 | 8.31 | 8.31 | 8.72 | 10.46 |
| 1n7f | 86 | 8 | 4.18 | 7.13 | 7.81 | 9.15 | 4.22 | 7.07 | 12.76 | 12.42 | 4.72 | 6.2 | 10.91 | 13.1 | 4.18 | 6.2 | 7.81 | 9.15 |
| 1nln | 203 | 11 | 3.34 | 3.62 | 3.97 | 5.23 | 5.57 | 7.07 | 8.61 | 12.81 | 3.07 | 3.7 | 3.74 | 12.51 | 3.07 | 3.62 | 3.74 | 5.23 |
| 1nq7 | 244 | 10 | 2.85 | 2.85 | 2.85 | 8.31 | 1.03 | 1.03 | 1.11 | 4.66 | 1.17 | 1.17 | 2.47 | 2.73 | 1.03 | 1.03 | 1.11 | 2.73 |
| 1ntv | 152 | 10 | 3.14 | 3.14 | 5.92 | 10 | 2.89 | 2.89 | 3.71 | 5.28 | 3.6 | 4.3 | 5.39 | 14.3 | 2.89 | 2.89 | 3.71 | 5.28 |
| 1nvr | 264 | 5 | 5.24 | 6.2 | 9.11 | 8.68 | 2.83 | 2.83 | 6.39 | 7.5 | 1.68 | 1.68 | 4.06 | 4.15 | 1.68 | 1.68 | 4.06 | 4.15 |
| 1nx1 | 173 | 11 | 3.03 | 3.2 | 3.7 | 4.2 | 2.82 | 2.98 | 3.26 | 4.68 | 2.92 | 2.94 | 3.33 | 4.95 | 2.82 | 2.94 | 3.26 | 4.2 |

| receptor pdb id | receptor length (AA) | peptide length (AA) | Run 1 | | | | Run 2 | | | | Run 3 | | | | best from all runs | | | |
|---|---|---|---|---|---|---|---|---|---|---|---|---|---|---|---|---|---|---|
| | | | all | top 1k | top 100 | top 10 | all | top 1k | top 100 | top 10 | all | top 1k | top 100 | top 10 | all | top 1k | top 100 | top 10 |
| 1oai | 59 | 9 | 3.81 | 4.86 | 5.27 | 6.84 | 3.04 | 3.56 | 5.54 | 6.68 | 3.23 | 4.75 | 6.18 | 6.53 | 3.04 | 3.56 | 5.27 | 6.53 |
| 1ou8 | 106 | 8 | 4.17 | 4.17 | 5.93 | 9.04 | 4.86 | 7.38 | 9.21 | 10.79 | 2.98 | 3.97 | 5.65 | 5.56 | 2.98 | 3.97 | 5.65 | 5.56 |
| 1pz5 | 435 | 8 | 4.58 | 4.58 | 5.8 | 5.63 | 4.69 | 5.18 | 5.32 | 5.22 | 5.02 | 5.97 | 7.73 | 8.22 | 4.58 | 4.58 | 5.32 | 5.22 |
| 1qkz | 219 | 10 | 5 | 6.45 | 10.31 | 15.24 | 6.71 | 7.72 | 11.97 | 13.27 | 6.96 | 6.96 | 6.96 | 10.69 | 5 | 6.45 | 6.96 | 10.69 |
| 1rxz | 245 | 11 | 6.76 | 6.76 | 11.68 | 13.38 | 4.46 | 5.22 | 8.5 | 11.28 | 3.54 | 3.9 | 6.16 | 6.11 | 3.54 | 3.9 | 6.16 | 6.11 |
| 1se0 | 97 | 7 | 6.16 | 7.78 | 11.06 | 17.95 | 4.05 | 4.05 | 9.08 | 7.55 | 5.34 | 7.02 | 9 | 9.06 | 4.05 | 4.05 | 9 | 7.55 |
| 1sfi | 223 | 14 | 7.8 | 7.8 | 10.63 | 13.04 | 7.18 | 7.18 | 7.36 | 7.77 | 6.74 | 8.53 | 8.78 | 11.51 | 6.74 | 7.18 | 7.36 | 7.77 |
| 1ssh | 60 | 11 | 3 | 4.79 | 5.35 | 7.05 | 4.19 | 4.67 | 5.44 | 5.41 | 3.74 | 4.32 | 5.33 | 5.68 | 3 | 4.32 | 5.33 | 5.41 |
| 1svz | 232 | 6 | 2.98 | 2.98 | 3.82 | 6.82 | 2.17 | 2.17 | 3.77 | 5.13 | 2.8 | 4.43 | 5.05 | 5.19 | 2.17 | 2.17 | 3.77 | 5.13 |
| 1t4f | 88 | 9 | 2.59 | 3.1 | 3.82 | 4.58 | 2.81 | 2.88 | 2.93 | 3.02 | 2.78 | 2.79 | 3.43 | 4.23 | 2.59 | 2.79 | 2.93 | 3.02 |
| 1t7r | 250 | 10 | 1.94 | 2.15 | 2.52 | 3.35 | 1.61 | 2.05 | 2.05 | 2.13 | 1.73 | 1.92 | 2.92 | 1.88 | 1.61 | 1.92 | 2.05 | 1.88 |
| 1tp5 | 115 | 6 | 4.28 | 7.19 | 9.28 | 8.47 | 1.68 | 1.73 | 2.64 | 3.57 | 1.15 | 1.45 | 2.46 | 3.76 | 1.15 | 1.45 | 2.46 | 3.57 |
| 1tw6 | 95 | 6 | 3.31 | 3.93 | 10.22 | 8.47 | 3.27 | 5.75 | 10.99 | 17.83 | 4.33 | 7.61 | 11.44 | 7.22 | 3.27 | 3.93 | 10.22 | 7.22 |
| 1u00 | 227 | 9 | 8.89 | 8.89 | 10.59 | 11.43 | 9.67 | 10.85 | 11.07 | 11.86 | 8.95 | 9.28 | 9.28 | 11.21 | 8.89 | 8.89 | 9.28 | 11.21 |
| 1u8i | 441 | 7 | 5.85 | 13.62 | 19.63 | 19.99 | 6.5 | 9.66 | 13.42 | 13.84 | 5.75 | 9.79 | 10.1 | 16.59 | 5.75 | 9.66 | 10.1 | 13.84 |
| 1u9l | 138 | 7 | 10.81 | 10.81 | 11.61 | 12.46 | 8.77 | 8.77 | 8.96 | 9.25 | 10.06 | 10.06 | 10.76 | 10.64 | 8.77 | 8.77 | 8.96 | 9.25 |
| 1uj0 | 58 | 9 | 3.44 | 3.53 | 4.29 | 5.19 | 3.49 | 3.49 | 4.57 | 4.83 | 4.1 | 4.71 | 4.9 | 5.37 | 3.44 | 3.49 | 4.29 | 4.83 |
| 1vzq | 250 | 6 | 2.98 | 4.02 | 6.61 | 10.51 | 2.76 | 4.5 | 5.03 | 6.71 | 3.73 | 5.85 | 6.35 | 7.21 | 2.76 | 4.02 | 5.03 | 6.71 |
| 1w9e | 164 | 5 | 3.77 | 6.17 | 11.09 | 17.41 | 5.4 | 5.76 | 9.04 | 16.41 | 4.19 | 13.05 | 16.34 | 17.86 | 3.77 | 5.76 | 9.04 | 16.41 |
| 1x2r | 290 | 9 | 4.95 | 5.02 | 5.04 | 5.1 | 2.89 | 4.85 | 6.54 | 7.13 | 3.89 | 5 | 5.78 | 7.1 | 2.89 | 4.85 | 5.04 | 5.1 |
| 1xoc | 504 | 9 | 19.49 | 19.49 | 20.94 | 21.66 | 17.64 | 17.97 | 18.06 | 20.33 | 18.15 | 18.15 | 18.15 | 19.29 | 17.64 | 17.97 | 18.06 | 19.29 |
| 1ymt | 235 | 10 | 2.98 | 2.99 | 4.12 | 5.98 | 2.54 | 2.77 | 2.89 | 3.28 | 2.79 | 3.53 | 3.72 | 4.54 | 2.54 | 2.77 | 2.89 | 3.28 |
| 1yph | 228 | 10 | 4.23 | 5.19 | 5.19 | 7.27 | 4.03 | 4.03 | 4.66 | 4.63 | 3.27 | 12.4 | 13.23 | 13.44 | 3.27 | 4.03 | 4.66 | 4.63 |
| 1yuc | 240 | 14 | 3.29 | 5.35 | 6.96 | 7.92 | 2.89 | 3.68 | 5.25 | 6.13 | 4.83 | 8.22 | 12.26 | 13 | 2.89 | 3.68 | 5.25 | 6.13 |
| 1ywo | 55 | 10 | 2.97 | 5.09 | 5.98 | 7.62 | 3.95 | 4.25 | 6.8 | 8.07 | 4.21 | 4.6 | 6.01 | 6.46 | 2.97 | 4.25 | 5.98 | 6.46 |
| 1z9o | 238 | 9 | 4.61 | 6.68 | 7.19 | 8.37 | 5.18 | 6.44 | 6.83 | 8.28 | 5.12 | 6.03 | 7.31 | 8.63 | 4.61 | 6.03 | 6.83 | 8.28 |
| 1zuk | 130 | 11 | 2.93 | 3.48 | 4.53 | 5.19 | 3.18 | 3.18 | 4.58 | 4.43 | 3.2 | 3.38 | 3.64 | 4.46 | 2.93 | 3.18 | 3.64 | 4.43 |
| 2a3i | 253 | 12 | 3.59 | 4.14 | 5.31 | 5.25 | 1.7 | 2.56 | 3.54 | 4.02 | 1.68 | 2.1 | 2.1 | 5.49 | 1.68 | 2.1 | 2.1 | 4.02 |
| 2ai4 | 111 | 8 | 5.94 | 8.48 | 8.51 | 9.49 | 9.6 | 11.5 | 14.81 | 15.92 | 7.93 | 11 | 13.87 | 14.01 | 5.94 | 8.48 | 8.51 | 9.49 |
| 2ak5 | 64 | 8 | 3.59 | 3.6 | 4.79 | 6.69 | 4.29 | 4.29 | 6.56 | 6.57 | 3.57 | 3.57 | 5.23 | 5.38 | 3.57 | 3.57 | 4.79 | 5.38 |
| 2b1z | 238 | 9 | 0.91 | 1.15 | 1.31 | 1.5 | 2.28 | 2.51 | 2.52 | 3.46 | 0.82 | 1.05 | 1.18 | 1.16 | 0.82 | 1.05 | 1.18 | 1.16 |
| 2b9h | 337 | 12 | 5.19 | 9.46 | 9.7 | 10.71 | 2.57 | 2.9 | 3.11 | 2.96 | 4.53 | 4.53 | 14 | 21.87 | 2.57 | 2.9 | 3.11 | 2.96 |
| 2bba | 185 | 14 | 4.08 | 4.79 | 6.87 | 7.1 | 3.53 | 3.65 | 5.17 | 5.26 | 3.91 | 4.58 | 5.41 | 5.57 | 3.53 | 3.65 | 5.17 | 5.26 |
| 2c3i | 266 | 8 | 6.09 | 7.29 | 8.93 | 11.16 | 5.12 | 6.69 | 8.61 | 11.37 | 3.16 | 6.05 | 9.94 | 11.43 | 3.16 | 6.05 | 8.61 | 11.16 |
| 2cch | 256 | 12 | 6.04 | 7.77 | 10.11 | 12.51 | 4.38 | 5.26 | 11.49 | 11.47 | 3.36 | 7.15 | 11.11 | 26.01 | 3.36 | 5.26 | 10.11 | 11.47 |

| receptor pdb id | receptor length (AA) | peptide length (AA) | Run 1 | | | | Run 2 | | | | Run 3 | | | | best from all runs | | | |
|---|---|---|---|---|---|---|---|---|---|---|---|---|---|---|---|---|---|---|
| | | | all | top 1k | top 100 | top 10 | all | top 1k | top 100 | top 10 | all | top 1k | top 100 | top 10 | all | top 1k | top 100 | top 10 |
| 2d0n | 56 | 9 | 3.41 | 3.66 | 4.42 | 6.02 | 3.27 | 3.67 | 4.23 | 4.68 | 2.64 | 2.64 | 4.22 | 5.64 | 2.64 | 2.64 | 4.22 | 4.68 |
| 2d5w | 602 | 5 | 18.64 | 19.17 | 19.48 | 21.23 | 17.99 | 18.59 | 20.08 | 20.39 | 18.52 | 18.52 | 19.65 | 20.13 | 17.99 | 18.52 | 19.48 | 20.13 |
| 2ds8 | 41 | 6 | 4.97 | 4.97 | 10.05 | 9.46 | 3.25 | 3.25 | 8.87 | 9.77 | 2.15 | 2.15 | 5.45 | 9.42 | 2.15 | 2.15 | 5.45 | 9.42 |
| 2dze | 320 | 6 | 2.39 | 2.71 | 2.71 | 3.13 | 2.82 | 2.97 | 3.23 | 3.59 | 7.08 | 7.45 | 7.8 | 8.58 | 2.39 | 2.71 | 2.71 | 3.13 |
| 2fgr | 332 | 8 | 6.29 | 13.42 | 14.19 | 15.87 | 9.8 | 13.72 | 14.8 | 16.07 | 4.96 | 13.75 | 13.75 | 16.49 | 4.96 | 13.42 | 13.75 | 15.87 |
| 2fmf | 128 | 13 | 2.87 | 2.87 | 7.07 | 7.5 | 4.46 | 4.64 | 6.09 | 7.18 | 4.77 | 5.14 | 5.92 | 6.43 | 2.87 | 2.87 | 5.92 | 6.43 |
| 2fnt | 198 | 7 | 0.76 | 0.8 | 0.8 | 1.12 | 4.94 | 4.94 | 4.98 | 5.68 | 8.72 | 8.72 | 8.72 | 10.73 | 0.76 | 0.8 | 0.8 | 1.12 |
| 2foj | 137 | 7 | 3.29 | 3.73 | 3.94 | 5.97 | 2.65 | 2.92 | 4.34 | 4.4 | 5.38 | 5.38 | 8.06 | 13.62 | 2.65 | 2.92 | 3.94 | 4.4 |
| 2fvj | 258 | 10 | 1.62 | 1.82 | 1.82 | 2.43 | 1.46 | 1.46 | 1.46 | 3.49 | 1.73 | 2.86 | 3.31 | 3.66 | 1.46 | 1.46 | 1.46 | 2.43 |
| 2h9m | 304 | 5 | 1.79 | 1.8 | 1.8 | 4.42 | 3.1 | 3.1 | 3.43 | 4.96 | 1.36 | 3 | 3.22 | 4.86 | 1.36 | 1.8 | 1.8 | 4.42 |
| 2ho2 | 33 | 10 | 4.32 | 6.62 | 8.82 | 15.21 | 3.94 | 4.89 | 6.48 | 7.04 | 3.8 | 5.63 | 6.33 | 5.59 | 3.8 | 4.89 | 6.33 | 5.59 |
| 2hpl | 100 | 5 | 1.28 | 2.66 | 3.45 | 3.94 | 1.99 | 2.34 | 4.06 | 5.22 | 1.76 | 2.08 | 4 | 5.16 | 1.28 | 2.08 | 3.45 | 3.94 |
| 2ipu | 442 | 7 | 5.04 | 5.5 | 6.47 | 8.75 | 7.4 | 7.57 | 9.74 | 17.13 | 5.42 | 7.49 | 11.56 | 10.82 | 5.04 | 5.5 | 6.47 | 8.75 |
| 2iv9 | 469 | 9 | 5.84 | 7.49 | 8.44 | 8.45 | 3.91 | 4.1 | 7.82 | 8.3 | 3.06 | 4.9 | 6.59 | 6.73 | 3.06 | 4.1 | 6.59 | 6.73 |
| 2j6f | 57 | 8 | 2.96 | 5.25 | 11.17 | 11.37 | 3.33 | 3.49 | 3.95 | 5.29 | 2.98 | 3.73 | 4.42 | 5.37 | 2.96 | 3.49 | 3.95 | 5.29 |
| 2jam | 279 | 6 | 4.98 | 7.86 | 12.7 | 15.99 | 4.34 | 7.51 | 11.81 | 17.2 | 6.16 | 8.41 | 12.8 | 13.62 | 4.34 | 7.51 | 11.81 | 13.62 |
| 2o02 | 224 | 14 | 3.99 | 5.54 | 6.41 | 6.68 | 3.97 | 4.08 | 4.86 | 4.92 | 3.54 | 3.71 | 4 | 5.19 | 3.54 | 3.71 | 4 | 4.92 |
| 2o4j | 240 | 12 | 1.49 | 2.26 | 2.41 | 2.81 | 1.54 | 2.02 | 2.69 | 2.95 | 6.07 | 7.81 | 11.78 | 12.76 | 1.49 | 2.02 | 2.41 | 2.81 |
| 2o9v | 67 | 10 | 3.64 | 3.85 | 5.4 | 6.6 | 3.76 | 4.35 | 5.36 | 9.44 | 3.38 | 3.73 | 4.6 | 5.39 | 3.38 | 3.73 | 4.6 | 5.39 |
| 2otu | 233 | 11 | 4.33 | 4.33 | 16.41 | 15.01 | 3.15 | 4.24 | 6.09 | 8.79 | 4.57 | 7.24 | 16.38 | 21.54 | 3.15 | 4.24 | 6.09 | 8.79 |
| 2p0w | 319 | 15 | 5.88 | 7.98 | 7.98 | 14.35 | 6.05 | 14.01 | 14.57 | 15.25 | 6.26 | 6.26 | 6.38 | 15.07 | 5.88 | 6.26 | 6.38 | 14.35 |
| 2p1k | 87 | 11 | 3.52 | 4.76 | 5.27 | 5.38 | 6.33 | 9.82 | 10 | 10.33 | 2.47 | 2.47 | 3.06 | 4.08 | 2.47 | 2.47 | 3.06 | 4.08 |
| 2p1t | 211 | 10 | 0.97 | 1.34 | 1.35 | 1.37 | 1.38 | 1.61 | 2.15 | 2.79 | 0.9 | 0.9 | 1.3 | 1.44 | 0.9 | 0.9 | 1.3 | 1.37 |
| 2p54 | 267 | 12 | 3.35 | 3.35 | 3.54 | 3.96 | 3.07 | 3.07 | 3.61 | 3.85 | 1.66 | 2.25 | 2.61 | 2.02 | 1.66 | 2.25 | 2.61 | 2.02 |
| 2puy | 60 | 10 | 6.13 | 7.84 | 18.61 | 21.21 | 6.3 | 6.3 | 10.36 | 10.35 | 3.63 | 3.63 | 8.7 | 12.44 | 3.63 | 3.63 | 8.7 | 10.35 |
| 2pv2 | 206 | 12 | 2.27 | 2.27 | 3.6 | 3.77 | 3.26 | 3.81 | 4.73 | 5.82 | 1.72 | 1.72 | 1.72 | 4.43 | 1.72 | 1.72 | 1.72 | 3.77 |
| 2qos | 173 | 11 | 3.11 | 4.16 | 6.57 | 6.88 | 3.53 | 4.83 | 5.87 | 6.79 | 3.17 | 3.94 | 4.65 | 5.76 | 3.11 | 3.94 | 4.65 | 5.76 |
| 2r7g | 337 | 10 | 1.47 | 1.62 | 1.62 | 3.33 | 2.34 | 2.85 | 3.53 | 4.97 | 2.94 | 3.7 | 4.03 | 5.09 | 1.47 | 1.62 | 1.62 | 3.33 |
| 2v3s | 96 | 6 | 2.42 | 2.42 | 3.48 | 8.89 | 1.83 | 1.83 | 5.01 | 7.24 | 1.05 | 1.09 | 1.74 | 2 | 1.05 | 1.09 | 1.74 | 2 |
| 2vj0 | 246 | 8 | 2.09 | 2.96 | 4.12 | 3.91 | 3.83 | 8 | 21.4 | 21.43 | 3.57 | 5.07 | 6.5 | 7.75 | 2.09 | 2.96 | 4.12 | 3.91 |
| 2zjd | 121 | 10 | 2.35 | 2.69 | 2.79 | 4.6 | 1.77 | 1.83 | 1.91 | 3.78 | 1.57 | 1.58 | 2.27 | 3.7 | 1.57 | 1.58 | 1.91 | 3.7 |
| 3bfq | 132 | 15 | 10.2 | 11.53 | 14.22 | 13.48 | 1.24 | 1.41 | 1.83 | 2.58 | 11.81 | 18.52 | 21.98 | 23.03 | 1.24 | 1.41 | 1.83 | 2.58 |
| 3bu3 | 297 | 14 | 6.87 | 7.06 | 7.71 | 8.86 | 5.77 | 5.77 | 10.23 | 11.07 | 6.12 | 10.01 | 10.01 | 10.42 | 5.77 | 5.77 | 7.71 | 8.86 |
| 3bwa | 276 | 8 | 2.42 | 2.75 | 3.17 | 3.94 | 2.39 | 2.51 | 2.98 | 3.2 | 1.83 | 2.7 | 3.19 | 3.65 | 1.83 | 2.51 | 2.98 | 3.2 |
| 3cvp | 279 | 6 | 4.52 | 4.67 | 8.47 | 10.14 | 3.07 | 4.97 | 5.68 | 7.59 | 3.08 | 4.25 | 5.49 | 8.02 | 3.07 | 4.25 | 5.49 | 7.59 |

| receptor pdb id | receptor length (AA) | peptide length (AA) | Run 1 | | | | Run 2 | | | | Run 3 | | | | best from all runs | | | |
|---|---|---|---|---|---|---|---|---|---|---|---|---|---|---|---|---|---|---|
| | | | all | top 1k | top 100 | top 10 | all | top 1k | top 100 | top 10 | all | top 1k | top 100 | top 10 | all | top 1k | top 100 | top 10 |
| 3d1e | 366 | 6 | 4.39 | 6.59 | 8.18 | 18.82 | 3.44 | 3.44 | 9.96 | 8.93 | 2.57 | 2.57 | 9.88 | 12.21 | 2.57 | 2.57 | 8.18 | 8.93 |
| 3d9t | 95 | 6 | 3.56 | 4.34 | 7.25 | 10.06 | 3.12 | 4.28 | 7.25 | 9.16 | 3.02 | 3.91 | 11.05 | 15.91 | 3.02 | 3.91 | 7.25 | 9.16 |
| **MEAN** | **204.81** | **9.17** | **4.41** | **5.53** | **7.32** | **9.11** | **4.38** | **5.32** | **7.13** | **8.52** | **4.38** | **5.71** | **7.49** | **9.26** | **3.61** | **4.32** | **5.62** | **6.90** |

Table S2. CABS-dock performance for 68 unbound cases (listed in rows) in 3 independent prediction runs (shown in separate columns). The table shows the lowest ligand-RMSD values (calculated on the peptide only after superimposition of the receptor structures) among: all 10,000 models (all), top 1000 models (top 1k, selected during filtering and clustering procedure), top 100 models (top 100, selected through further clustering) and top 10 final models (top 10). Last column shows the lowest RMSD values obtained in three prediction runs.

| pdb code | receptor length (AA) | peptide length (AA) | run 1 | | | | run 2 | | | | run 3 | | | | best from 3 runs | | | |
|---|---|---|---|---|---|---|---|---|---|---|---|---|---|---|---|---|---|---|
| | | | all | top 1k | top 100 | top 10 | all | top 1k | top 100 | top 10 | all | top 1k | top 100 | top 10 | all | top 1k | top 100 | top 10 |
| 1alv | 173 | 11 | 1.99 | 2.55 | 3.6 | 3.32 | 2.83 | 3.24 | 3.73 | 4.87 | 1.68 | 2.12 | 3.2 | 2.47 | 1.68 | 2.12 | 3.2 | 2.47 |
| 1b9k | 246 | 8 | 3.38 | 6.89 | 9.74 | 9.95 | 4.07 | 10.06 | 14.05 | 22.2 | 4.74 | 6.18 | 10.07 | 10.59 | 3.38 | 6.18 | 9.74 | 9.95 |
| 1czz | 168 | 7 | 2.56 | 2.56 | 3.62 | 3.81 | 2.03 | 2.33 | 2.58 | 3.21 | 2.06 | 2.83 | 3.15 | 5.68 | 2.03 | 2.33 | 2.58 | 3.21 |
| 1d1z | 104 | 11 | 5.26 | 5.26 | 7.09 | 7.48 | 5.11 | 5.59 | 6.85 | 10.01 | 5.17 | 5.33 | 7.21 | 8.02 | 5.11 | 5.26 | 6.85 | 7.48 |
| 1eg3 | 260 | 13 | 6.17 | 10.8 | 14.45 | 16.05 | 6.67 | 8.44 | 12.43 | 22.77 | 7.47 | 9.46 | 14.72 | 15.02 | 6.17 | 8.44 | 12.43 | 15.02 |
| 1go5 | 59 | 9 | 3.38 | 3.92 | 4.92 | 5.22 | 2.89 | 3.4 | 3.65 | 4.66 | 4.17 | 4.67 | 11.15 | 12.66 | 2.89 | 3.4 | 3.65 | 4.66 |
| 1gy7 | 122 | 5 | 3.73 | 7.64 | 13.05 | 13.73 | 6.01 | 7.96 | 12.23 | 15.15 | 4.46 | 4.59 | 12.19 | 13.71 | 3.73 | 4.59 | 12.19 | 13.71 |
| 1h1r | 256 | 12 | 4.37 | 4.92 | 7.09 | 6.55 | 4.42 | 5.04 | 10.24 | 10.18 | 3.86 | 6.12 | 7.99 | 8.03 | 3.86 | 4.92 | 7.09 | 6.55 |
| 1i2h | 104 | 6 | 2.71 | 3.33 | 3.95 | 4.31 | 3.77 | 3.77 | 8.36 | 10.09 | 3.29 | 3.33 | 4.64 | 4.63 | 2.71 | 3.33 | 3.95 | 4.31 |
| 1i7g | 267 | 12 | 1.68 | 2.01 | 2.87 | 7.9 | 2.26 | 2.47 | 2.78 | 3.38 | 1.75 | 1.83 | 1.83 | 1.87 | 1.68 | 1.83 | 1.83 | 1.87 |
| 1ie9 | 240 | 12 | 6.93 | 11.86 | 13.99 | 23.44 | 4.85 | 7.75 | 10.22 | 12.85 | 6.88 | 10.98 | 12.36 | 15.34 | 4.85 | 7.75 | 10.22 | 12.85 |
| 1jbe | 128 | 13 | 4.84 | 5.44 | 5.75 | 6.08 | 4.52 | 5.76 | 5.92 | 6.44 | 4.37 | 4.58 | 7.97 | 10.79 | 4.37 | 4.58 | 5.75 | 6.08 |
| 1jwf | 140 | 5 | 2.17 | 3.89 | 4.46 | 4.7 | 2.52 | 2.52 | 4.94 | 5.71 | 2.35 | 2.46 | 4.24 | 5.96 | 2.17 | 2.46 | 4.24 | 4.7 |
| 1jwt | 250 | 6 | 9.44 | 14.28 | 17.25 | 17.23 | 8.68 | 11.8 | 14.42 | 15.36 | 9.8 | 14.08 | 17.4 | 19.47 | 8.68 | 11.8 | 14.42 | 15.36 |
| 1lf7 | 173 | 11 | 5.68 | 7.47 | 7.6 | 7.74 | 4.1 | 4.97 | 4.98 | 4.83 | 4.89 | 5.53 | 6.87 | 7.59 | 4.1 | 4.97 | 4.98 | 4.83 |
| 1lvb | 229 | 7 | 5.73 | 7.41 | 13.21 | 18.56 | 5.73 | 13.74 | 14.39 | 18.3 | 7.12 | 7.12 | 15.6 | 17.56 | 5.73 | 7.12 | 13.21 | 17.56 |
| 1m7d | 435 | 8 | 5.61 | 6.35 | 7.4 | 13.37 | 5.29 | 5.84 | 6.78 | 13.62 | 3.94 | 4.67 | 5.5 | 5.89 | 3.94 | 4.67 | 5.5 | 5.89 |
| 1n7e | 86 | 8 | 4.79 | 4.79 | 8.03 | 11.97 | 4.08 | 8.01 | 11.44 | 11.26 | 3.79 | 6.19 | 10.49 | 16.36 | 3.79 | 4.79 | 8.03 | 11.26 |
| 1n83 | 244 | 10 | 1.12 | 1.13 | 1.13 | 2.39 | 2.73 | 2.73 | 3.74 | 3.84 | 1.48 | 3.56 | 8.3 | 8.56 | 1.12 | 1.13 | 1.13 | 2.39 |
| 1oew | 330 | 8 | 7.92 | 8.65 | 9.87 | 10.63 | 5.72 | 8.05 | 8.09 | 9.18 | 6.58 | 6.58 | 7.84 | 9 | 5.72 | 6.58 | 7.84 | 9 |
| 1oot | 60 | 11 | 3.43 | 4.62 | 5.97 | 6.97 | 4.27 | 4.97 | 6.1 | 6.9 | 3.97 | 4.62 | 5.73 | 7.12 | 3.43 | 4.62 | 5.73 | 6.9 |
| 1ou9 | 106 | 8 | 3.58 | 4.21 | 4.97 | 6.78 | 3.64 | 3.64 | 4.93 | 11.23 | 3.25 | 4.56 | 4.81 | 5.18 | 3.25 | 3.64 | 4.81 | 5.18 |
| 1pyw | 369 | 15 | 3.62 | 3.65 | 3.79 | 4.68 | 4.62 | 5.23 | 5.54 | 5.8 | 3.93 | 5.5 | 5.55 | 6.33 | 3.62 | 3.65 | 3.79 | 4.68 |
| 1qbh | 105 | 8 | 4.72 | 7.01 | 9.44 | 13.51 | 5.44 | 6 | 8.96 | 12.8 | 6.09 | 6.09 | 8.51 | 10.44 | 4.72 | 6 | 8.51 | 10.44 |
| 1r6j | 164 | 5 | 2.52 | 2.52 | 8.26 | 9.3 | 3.21 | 3.23 | 5.43 | 8.31 | 3.2 | 3.72 | 4.94 | 8.47 | 2.52 | 2.52 | 4.94 | 8.31 |
| 1rwz | 245 | 11 | 6.17 | 10.34 | 11.2 | 23.02 | 4.71 | 4.71 | 8.11 | 19.96 | 4.73 | 4.89 | 6.67 | 7.61 | 4.71 | 4.71 | 6.67 | 7.61 |
| 1tq3 | 115 | 6 | 1.4 | 1.4 | 3.57 | 6.44 | 3.58 | 5.63 | 7.34 | 7.39 | 2.62 | 2.62 | 4.7 | 9.68 | 1.4 | 1.4 | 3.57 | 6.44 |
| 1um5 | 431 | 8 | 3.05 | 3.05 | 3.34 | 3.93 | 3.54 | 3.54 | 5.67 | 4.78 | 4.32 | 4.32 | 5.89 | 7.93 | 3.05 | 3.05 | 3.34 | 3.93 |
| 1utn | 223 | 14 | 7.74 | 8.46 | 8.73 | 12.33 | 6.95 | 7.66 | 8.69 | 10.88 | 7.27 | 7.91 | 9.31 | 10.01 | 6.95 | 7.66 | 8.69 | 10.01 |

| pdb code | receptor length (AA) | peptide length (AA) | run 1 | | | | run 2 | | | | run 3 | | | | best from 3 runs | | | |
|---|---|---|---|---|---|---|---|---|---|---|---|---|---|---|---|---|---|---|
| | | | all | top 1k | top 100 | top 10 | all | top 1k | top 100 | top 10 | all | top 1k | top 100 | top 10 | all | top 1k | top 100 | top 10 |
| 1v49 | 121 | 10 | 2.69 | 2.69 | 2.91 | 4.02 | 2.19 | 2.56 | 2.73 | 3.45 | 2.89 | 2.89 | 4.23 | 5.5 | 2.19 | 2.56 | 2.73 | 3.45 |
| 1x2j | 290 | 9 | 3.92 | 5.08 | 5.42 | 6 | 3.44 | 3.68 | 5.17 | 5.89 | 3.57 | 4.92 | 5.81 | 6.54 | 3.44 | 3.68 | 5.17 | 5.89 |
| 1y0m | 55 | 10 | 3.92 | 4.97 | 5.73 | 7.03 | 3.53 | 4.83 | 5.45 | 6.21 | 4 | 5.24 | 5.84 | 8.13 | 3.53 | 4.83 | 5.45 | 6.21 |
| 1yej | 442 | 7 | 4.67 | 5.2 | 10.97 | 19.17 | 5.31 | 6.37 | 8.26 | 12.12 | 4.05 | 5.94 | 8.05 | 7.67 | 4.05 | 5.2 | 8.05 | 7.67 |
| 1z1m | 88 | 9 | 2.76 | 3.46 | 3.72 | 4.68 | 3.31 | 3.48 | 5.26 | 4.54 | 3.39 | 3.52 | 3.75 | 3.62 | 2.76 | 3.46 | 3.72 | 3.62 |
| 1z9l | 238 | 9 | 4.24 | 5.32 | 6.68 | 8.65 | 4.18 | 4.87 | 5.84 | 6.91 | 3.8 | 4.84 | 4.91 | 5.08 | 3.8 | 4.84 | 4.91 | 5.08 |
| 2aa2 | 253 | 12 | 1.84 | 2 | 2.95 | 3.12 | 2.32 | 2.52 | 2.54 | 2.95 | 1.73 | 2.39 | 2.85 | 4.12 | 1.73 | 2 | 2.54 | 2.95 |
| 2abx | 74 | 13 | 6.14 | 7.58 | 10.07 | 10.19 | 6.5 | 7.22 | 7.96 | 8.76 | 8.02 | 9.53 | 10.21 | 11.18 | 6.14 | 7.22 | 7.96 | 8.76 |
| 2alf | 164 | 6 | 1.74 | 2.42 | 2.7 | 4.1 | 2.17 | 2.29 | 3.38 | 3.46 | 1.8 | 1.81 | 2.75 | 2.55 | 1.74 | 1.81 | 2.7 | 2.55 |
| 2am9 | 250 | 10 | 1.52 | 1.58 | 1.69 | 2.14 | 1.57 | 1.93 | 2.6 | 2.22 | 1.21 | 1.76 | 2.36 | 2.66 | 1.21 | 1.58 | 1.69 | 2.14 |
| 2b9f | 337 | 12 | 2.44 | 2.66 | 3.52 | 3.86 | 3.85 | 3.85 | 6.1 | 12.32 | 2.33 | 3.62 | 4.09 | 4.16 | 2.33 | 2.66 | 3.52 | 3.86 |
| 2bz6 | 239 | 15 | 8.03 | 16.66 | 18.28 | 22.88 | 7.76 | 16.35 | 21.5 | 23.07 | 6 | 9.88 | 10.72 | 10.51 | 6 | 9.88 | 10.72 | 10.51 |
| 2ds7 | 41 | 6 | 5.64 | 5.64 | 11.31 | 11.24 | 5.44 | 7.26 | 10.81 | 11.85 | 3.63 | 7.09 | 10.38 | 12.17 | 3.63 | 5.64 | 10.38 | 11.24 |
| 2dvj | 57 | 9 | 2.82 | 3.29 | 5.09 | 7.7 | 2.96 | 3.3 | 5.85 | 6.15 | 3.35 | 5.1 | 6.38 | 7.2 | 2.82 | 3.29 | 5.09 | 6.15 |
| 2e45 | 33 | 10 | 4.43 | 5.35 | 7.11 | 17.03 | 3.38 | 5.57 | 7.04 | 14.19 | 4.07 | 7.35 | 7.97 | 10.34 | 3.38 | 5.35 | 7.04 | 10.34 |
| 2f1w | 137 | 7 | 2.82 | 7.32 | 7.68 | 7.82 | 3.36 | 4.93 | 5.65 | 6.91 | 3.32 | 3.35 | 3.35 | 7.82 | 2.82 | 3.35 | 3.35 | 6.91 |
| 2fgq | 332 | 8 | 7.95 | 14.74 | 15.97 | 17.58 | 10.19 | 13.38 | 16.67 | 17.32 | 7.73 | 13.64 | 14.87 | 14.82 | 7.73 | 13.38 | 14.87 | 14.82 |
| 2g6f | 64 | 8 | 3.44 | 3.44 | 6.45 | 6.64 | 3.25 | 4.5 | 5.08 | 4.94 | 3.26 | 5.11 | 6.14 | 7.71 | 3.25 | 3.44 | 5.08 | 4.94 |
| 2h14 | 304 | 5 | 2.55 | 2.94 | 3.4 | 4.26 | 2.57 | 3.09 | 3.09 | 4.7 | 2.28 | 2.28 | 3.73 | 4.42 | 2.28 | 2.28 | 3.09 | 4.26 |
| 2h3l | 95 | 9 | 4.36 | 4.66 | 6.77 | 12.2 | 4.96 | 4.96 | 14.76 | 14.9 | 3.54 | 4.71 | 4.71 | 9.11 | 3.54 | 4.66 | 4.71 | 9.11 |
| 2hpj | 100 | 5 | 1.64 | 3.33 | 3.83 | 6.03 | 2.1 | 2.17 | 3.37 | 4.58 | 2.35 | 3.18 | 3.77 | 4.91 | 1.64 | 2.17 | 3.37 | 4.58 |
| 2hwq | 258 | 10 | 2.25 | 2.82 | 3.31 | 2.78 | 1.16 | 1.6 | 2.29 | 2.68 | 1.17 | 1.17 | 1.36 | 1.49 | 1.16 | 1.17 | 1.36 | 1.49 |
| 2i3i | 95 | 6 | 3.85 | 6.18 | 7.06 | 15.82 | 4.03 | 4.03 | 6.82 | 17.23 | 3.36 | 8.39 | 9.27 | 9.4 | 3.36 | 4.03 | 6.82 | 9.4 |
| 2iog | 238 | 9 | 9.98 | 10.51 | 14.36 | 23.02 | 8.77 | 10.68 | 13.32 | 21.29 | 9.46 | 10.35 | 15.22 | 20.05 | 8.77 | 10.35 | 13.32 | 20.05 |
| 2j2i | 266 | 8 | 3.06 | 4.03 | 10.45 | 11.4 | 3.75 | 6.46 | 8.73 | 11.68 | 3.92 | 6.3 | 9.38 | 11.1 | 3.06 | 4.03 | 8.73 | 11.1 |
| 2j6k | 57 | 8 | 3.27 | 3.8 | 4.07 | 7.3 | 2.98 | 3.39 | 3.39 | 5.62 | 2.77 | 3.06 | 3.34 | 4.03 | 2.77 | 3.06 | 3.34 | 4.03 |
| 2o9s | 67 | 10 | 3.24 | 3.86 | 5.01 | 6.96 | 3.23 | 4.87 | 5.68 | 6.36 | 3.92 | 4.36 | 5.48 | 6.85 | 3.23 | 3.86 | 5.01 | 6.36 |
| 2qbh | 95 | 6 | 5.11 | 5.99 | 9.29 | 10.33 | 5.21 | 6.29 | 14.34 | 14.67 | 5.03 | 5.03 | 7.98 | 12.62 | 5.03 | 5.03 | 7.98 | 10.33 |
| 2qhn | 264 | 5 | 4.9 | 6.27 | 7.28 | 9.16 | 3.74 | 3.87 | 7.1 | 5.22 | 4.62 | 5.12 | 8.55 | 9.79 | 3.74 | 3.87 | 7.1 | 5.22 |
| 2rtm | 120 | 6 | 4.44 | 5.13 | 6.41 | 6.93 | 5.28 | 5.57 | 5.96 | 6.79 | 5.09 | 5.25 | 6.2 | 6.49 | 4.44 | 5.13 | 5.96 | 6.49 |
| 2yql | 60 | 10 | 4.57 | 4.99 | 6.49 | 7.59 | 3.29 | 3.29 | 7.69 | 9.57 | 6.5 | 7.18 | 8.39 | 13.12 | 3.29 | 3.29 | 6.49 | 7.59 |
| 3d1g | 366 | 6 | 2.73 | 2.73 | 7.04 | 9 | 7.61 | 14.02 | 14.54 | 15.3 | 3.79 | 6.23 | 11.18 | 6.88 | 2.73 | 2.73 | 7.04 | 6.88 |
| 3ekk | 297 | 14 | 8.03 | 9.61 | 10.36 | 11.6 | 6.91 | 7.01 | 7.11 | 8.56 | 4.41 | 6.57 | 10.22 | 9.96 | 4.41 | 6.57 | 7.11 | 8.56 |
| 3hau | 198 | 7 | 8.66 | 10.03 | 10.04 | 11.73 | 6.24 | 6.24 | 6.8 | 7 | 8.11 | 8.34 | 8.36 | 9.08 | 6.24 | 6.24 | 6.8 | 7 |
| 3nsq | 211 | 10 | 3.19 | 4.38 | 6.44 | 10.29 | 2.82 | 3.91 | 4.37 | 8.73 | 2.4 | 2.64 | 3.76 | 5.76 | 2.4 | 2.64 | 3.76 | 5.76 |

| pdb code | receptor length (AA) | peptide length (AA) | run 1 | | | | run 2 | | | | run 3 | | | | best from 3 runs | | | |
|---|---|---|---|---|---|---|---|---|---|---|---|---|---|---|---|---|---|---|
| | | | all | top 1k | top 100 | top 10 | all | top 1k | top 100 | top 10 | all | top 1k | top 100 | top 10 | all | top 1k | top 100 | top 10 |
| 3pom | 337 | 10 | 3.59 | 3.59 | 3.59 | 6.69 | 3.99 | 4.61 | 5.26 | 8.79 | 2.17 | 2.17 | 3.89 | 6.42 | 2.17 | 2.17 | 3.59 | 6.42 |
| 3rdh | 224 | 14 | 3.79 | 4.13 | 6.16 | 6.95 | 3.58 | 3.93 | 4.24 | 6.28 | 3.88 | 3.91 | 4.88 | 4.96 | 3.58 | 3.91 | 4.24 | 4.96 |
| 3siq | 97 | 7 | 5.08 | 5.08 | 9.61 | 10.19 | 7.67 | 7.67 | 8.73 | 15.02 | 8.65 | 8.65 | 8.65 | 14.31 | 5.08 | 5.08 | 8.65 | 10.19 |
| 3tx7 | 240 | 14 | 4.01 | 5.77 | 11.11 | 12.75 | 4.14 | 5.07 | 5.28 | 8.25 | 3.22 | 3.69 | 8.73 | 12.59 | 3.22 | 3.69 | 5.28 | 8.25 |
| **MEAN** | **191.71** | **9.07** | **4.25** | **5.52** | **7.33** | **9.53** | **4.37** | **5.58** | **7.40** | **9.59** | **4.24** | **5.34** | **7.21** | **8.63** | **3.68** | **4.49** | **6.14** | **7.32** |

**Table S3. PDB codes of receptor pairs (in bound and unbound form).**

| bound form pdb code | unbound form pdb code |
|---|---|
| 1awr | 2alf |
| 1ce1 | 1um5 |
| 1cka | 2dvj |
| 1czy | 1czz |
| 1d4t | 1d1z |
| 1ddv | 1i2h |
| 1eg4 | 1eg3 |
| 1er8 | 1oew |
| 1gyb | 1gy7 |
| 1hc9 | 2abx |
| 1jbu | 2bz6 |
| 1jd5 | 1qbh |
| 1jwg | 1jwf |
| 1kl3 | 2rtm |
| 1klu | 1pyw |
| 1lvm | 1lvb |
| 1mfg | 2h3l |
| 1n7f | 1n7e |
| 1nq7 | 1n83 |
| 1nvr | 2qhn |
| 1nx1 | 1alv |
| 1oai | 1go5 |
| 1ou8 | 1ou9 |

| bound form pdb code | unbound form pdb code |
|---|---|
| 1pz5 | 1m7d |
| 1rxz | 1rwz |
| 1se0 | 3siq |
| 1sfi | 1utn |
| 1ssh | 1oot |
| 1t4f | 1z1m |
| 1t7r | 2am9 |
| 1tp5 | 1tq3 |
| 1tw6 | 2i3i |
| 1vzq | 1jwt |
| 1w9e | 1r6j |
| 1x2r | 1x2j |
| 1yuc | 3tx7 |
| 1ywo | 1y0m |
| 1z9o | 1z9l |
| 2a3i | 2aa2 |
| 2ak5 | 2g6f |
| 2b1z | 2iog |
| 2b9h | 2b9f |
| 2c3i | 2j2i |
| 2cch | 1h1r |
| 2ds8 | 2ds7 |
| 2fgr | 2fgq |
| 2fmf | 1jbe |
| 2fnt | 3hau |
| 2foj | 2f1w |
| 2fvj | 2hwq |
| 2h9m | 2h14 |
| 2ho2 | 2e45 |

| bound form pdb code | unbound form pdb code |
|---|---|
| 2hpl | 2hpj |
| 2ipu | 1yej |
| 2j6f | 2j6k |
| 2o02 | 3rdh |
| 2o4j | 1ie9 |
| 2o9v | 2o9s |
| 2p1t | 3nsq |
| 2p54 | 1i7g |
| 2puy | 2yql |
| 2qos | 1lf7 |
| 2r7g | 3pom |
| 2vj0 | 1b9k |
| 2zjd | 1v49 |
| 3bu3 | 3ekk |
| 3d1e | 3d1g |
| 3d9t | 2qbh |